\documentclass[pdftex,a4paper]{scrartcl}

\usepackage[ngerman,english]{babel}
\useshorthands{"}
\addto\extrasenglish{\languageshorthands{ngerman}}
\usepackage[utf8]{inputenc}
\usepackage[T1]{fontenc}
\usepackage{amsmath}
\usepackage{graphicx}
\usepackage{amssymb}
\usepackage{hyperref}
\usepackage{mathrsfs}
\usepackage{color}

\renewcommand{\phi}{\varphi}
\renewcommand{\theta}{\vartheta}

\title{{Gravitational Black Hole Hair from Event Horizon Supertranslations}}
\author{{\bf Artem Averin$^{\textrm{a,b}}$, Gia Dvali$^{\textrm{a,b,c}}$, Cesar Gomez$^{\textrm{d}}$, Dieter L\"ust$^{\textrm{a,b}}$}}

\begin{document}

\maketitle

\centerline{\it $^{\textrm{a}}$ Arnold--Sommerfeld--Center for Theoretical Physics,}
\centerline{\it Ludwig--Maximilians--Universit\"at, 80333 M\"unchen, Germany}
\medskip
\centerline{\it $^{\textrm{b}}$ Max--Planck--Institut f\"ur Physik,
Werner--Heisenberg--Institut,}
\centerline{\it 80805 M\"unchen, Germany}
\medskip
\centerline{\it $^{\textrm{c}}$ Center for Cosmology and Particle Physics,
Department of Physics, New York University}
\centerline{\it 4 Washington Place, New York, NY 10003, USA}
\medskip
\centerline{\it $^{\textrm{d}}$ Instituto de F\'{\i}sica Te\'orica UAM-CSIC, C-XVI,
Universidad Aut\'onoma de Madrid,}
\centerline{\it Cantoblanco, 28049 Madrid, Spain}

\vskip1cm
\abstract{{
We discuss BMS supertranslations both at null-infinity $BMS^-$ and on the horizon $BMS^{\cal H}$ for the case of the Schwarzschild black hole. We show that both kinds of supertranslations lead to infinetly many gapless physical excitations. On this basis we construct a quotient algebra $\cal A \equiv BMS^{\cal H}/BMS^-$ using suited superpositions of both kinds of transformations  which cannot be compensated by an ordinary BMS-supertranslation and therefore are intrinsically due to the presence of an event horizon. We show that transformations in $\cal A$ are physical and generate gapless excitations on the horizon that can account for the gravitational hair as well as for the black hole entropy. We identify the physics of these modes as associated with Bogolioubov-Goldstone modes due to quantum criticality. Classically 
the number of these gapless modes is infinite.  However, we show that due to quantum criticality the actual amount of information-carriers becomes finite and consistent with Bekenstein entropy. 
Although we only consider the case of Schwarzschild geometry, the arguments are extendable to arbitrary space-times containing event horizons.
}}

\begin{flushright}
{\small  MPP--2016--3},
{\small LMU--ASC 05/16}
\end{flushright}

\newpage

\setcounter{tocdepth}{2}
\tableofcontents
\break

\section{Introduction}

   For many decades a commonly accepted view was that a purely 
   gravitational hair of a black hole contains no information besides its mass 
 $M$ and an angular momentum.  This view is closely connected 
 to another statement that the Hawking spectrum \cite{Hawking} of black hole evaporation 
  is {\it exactly} thermal.  Both statements are based on semi-classical analysis and in order to understand when and how they go wrong, we have to define 
  various regimes rigorously. 
  
   In black hole physics the two important length-scales are: 
    The gravitational (Schwarzschild) radius,  $r_S \equiv 2G_NM$, and 
   the Planck length, $L_P^2 \equiv \hbar G_N$, where 
   $G_N$ and $\hbar$ are Newton's and Planck's constants respectively.     
     In this notations, the quantum gravitational coupling of gravitons of wavelength $R$ is given by 
    $\alpha \equiv {L_P^2 \over R^2}$. 
 
   Then, the semi-classical black hole regime corresponds to the following double-scaling 
   limit, 
\begin{equation}
 \hbar = {\rm finite}, ~ r_S = {\rm finite} ,~ G_N \rightarrow 0, ~ M \rightarrow \infty  
 \, .    
 \label{semilimit}
 \end{equation}    
 In this limit, any quantum back-reaction on the geometry can be safely ignored. 
 This is also apparent from the fact that in the above limit quantum gravitational coupling vanishes, $\alpha \rightarrow 0$.  
 
 The equation (\ref{semilimit}) describes the limit in which Hawking's famous calculation \cite{Hawking} has been done and it reveals that the spectrum of emitted particles is exactly thermal. However, 
regarding the question of a potential loss of information, this  analysis is inconclusive, because 
exactly in this limit the black hole is {\it eternal}. 
  In order to conclude whether there is any information puzzle, we need 
  to know what happens in the case of finite $M$ and $G_N$.  \\
  
   In \cite{Nportrait, gold} it was  concluded that gravitational hair of a black hole is capable of carrying its entire quantum information.  This conclusion was reached within a particular microscopic theory, but the result that we shall use, 
  are very general. Namely,  a black hole represents a system at the {\it quantum critical point},
   characterized by existence of many nearly-gapless collective modes, 
  the so-called Bogoliubov modes.  These modes are the carriers of the 
  black hole quantum information and they exhibit the following features: \\
 
{\it  1) Number of distinct modes scales as Bekenstein entropy, 
 \begin{equation}
   N \, = \, {r_S^2 \over L_P^2} \,;
 \end{equation} 
  
 2) The energy gap for exciting each mode scales as,  
 \begin{equation}
 \Delta E = {1 \over N} {\hbar \over r_S};
 \label{gap} 
 \end{equation}
 
 3)  The gravitational coupling, evaluated for wave-length $r_S$,  is  $\alpha = { 1\over N}$.   
 
 4) Deviations from thermal spectrum in Hawking radiation are 
 of order ${1 \over N}$.}\footnote{This property can be proven in very general terms \cite{nonthermal}.}
 \\
        
   Correspondingly, the hair-resolution time scales as 
   $t_{hair} \sim N r_S$ and is of the order of the black hole half life time. 
   Notice that, both in classical as well as in semi-classical limits, we have  $N \rightarrow \infty$.  Thus, according to \cite{Nportrait,gold} the purely gravitational hair of a 
   (semi)classical back hole in fact contains an {\it infinite}  amount of information. 
   However, the time-scale for resolving this information is also infinite.  
  This fact makes it clear that the exact thermality of Hawking spectrum in semi-classical theory is no ground for the information puzzle.  The 
 determining factor is the relation between the relative deviations from the thermal spectrum (i.e., the strength of black hole hair)  and the black hole life-time. 
  The deviations of order ${1 \over N}$ per emission time are fully enough both for storing the entire black hole information as well as for its consistent recovery \cite{Nportrait,gold} \\

  What is very important is to appreciate that the above features are completely 
  independent of a particular microscopic theory and represent 
  a parameterization of black hole properties obtained by superimposing 
  well-accepted facts, such as Bekenstein entropy \cite{Bent}, on quantum mechanics \cite{giamischa}. 
  
  This is the approach we shall adopt in the present work.
   We shall not make any particular assumption about the microscopic theory, but firmly demand that it exists and is compatible with the rules of quantum 
  mechanics. From the existence of Bekenstein entropy, it then directly follows 
   that black hole information-carrying degrees of freedom  must satisfy the properties {\it 1)-3)} \cite{giamischa}.  Indeed, in order to match the entropy, 
   the number of such degrees of freedom must scale as $N$.\footnote{Attributing all $2^N$ states to a single (or few) oscillators inevitably runs us into the problem.  In this case the only label that distinguishes states is energy. Then because of high degeneracy the resolution time would be exponentially large. So, information {\it must}  be stored in
   distinct degrees of freedom, in which case the states carry an extra label of 
  oscillator species and  everything is consistent} Moreover,  the energy needed for exciting individual modes, $\Delta E$, is fixed by the two constraints. First, the energy of bringing all the modes simultaneously into excited states should not exceed the energy of a 
 single Hawking quantum, i.e., $ N \Delta E \lesssim {\hbar \over r_S}$. Secondly, 
the minimal time-scale for resolving  the state of individual modes, from quantum uncertainty is  $t_{hair} \sim {\hbar \over \Delta E}$.  By consistency, the latter time-scale should not be much  longer than the black hole life-time.  These two constraints uniquely fix the gap as given by (\ref{gap}).  Finally, the relevant coupling is simply an ordinary gravitational coupling evaluated for the wavelength 
$r_S$, which gives $\alpha = {1 \over N}$.

   Notice, that since $\alpha N =1$ the quantum criticality 
comes out as an inevitable property, rather than an assumption. 
 Because of this universal property, despite the fact that we abstract from a particular microscopic picture, we shall keep the terminology applicable to 
 critical Bose-Einstein substances.  Namely, we shall continue to refer to  the 
information-carrying degrees of freedom as Bogoliubov modes and to the 
equations that these modes satisfy as Bogoliubov-de Gennes equation.  
 The classical geometric description is obtained as a double-scaling limit 
 in which $N \rightarrow \infty$ and  $\alpha N = 1$. \\ 
 
 The question that we shall investigate in the present paper is the following: 
What is the geometric description of the black hole quantum hair? 
 The answer to this question was already suggested in \cite{Dvali:2015rea} where it was 
 conjectured that the geometric description of Bogoliubov modes is in terms of Goldstone bosons of spontaneously broken BMS-type symmetry \cite{bms} acting on the black hole horizon.  

In the present paper we shall try to explicitly derive the broken symmetry transformations that are responsible for carrying black hole information and entropy.  As a possible candidate  we identify the part of BMS symmetries that act as the area-preserving diffeomorphisms on the black hole event horizon. The corresponding Goldstone bosons then are naturally identified  as the geometric limit 
of quantum Bogoliubov modes. The correspondence between quantum and geometric descriptions is summarized in table 1. \\

   \begin{table}
  \begin{center}
 \begin{tabular}{|l | l|}
  \hline
  \textbf{Quantum Picture: $N=$finite} & \textbf{Geometric Picture: $N \rightarrow \infty$}\\ 
   \hline
  Quantum Criticality &  Existence of the Horizon \\

  \hline
  Bogoliubov modes &  ${\cal A} \equiv BMS^{\cal H}/BMS^-$ Goldstones \\
  \hline
  Bogoliubov- de Gennes equation &  Equation for small metric deformations \\
  \hline
  Number of  modes $N=$finite & Number of modes  $N = \infty$ \\
  \hline
  Coupling  $\alpha = {1 \over N}=$finite & Coupling  $\alpha = {1 \over N} \rightarrow 0$ \\
  \hline
     \hline  Hair resolution time  $t_{hair} = NR$ & 
   Hair resolution time  $t_{hair} \rightarrow \infty$\\
       \hline  Energy gap $\Delta E = {1 \over N}{\hbar \over r_S}  $ &  Energy gap
    $\Delta E  \rightarrow  0  $ \\
       \hline
\end{tabular}
  \end{center}
\caption{Correspondence between the quantum and 
the classical-geometric pictures. \newline
The latter corresponds to the double scaling 
limit with $N \rightarrow \infty,~ \alpha N = 1$ }\label{BH}
\end{table}

  We shall now introduce the BMS side of the story. 
Over many years, asymptotic symmetries  of certain space-time geometries play an important role in general relativity. 
So far asymptotic symmetries are constructed at the boundaries of certain space-time geometries. In particular they play an important role on the AdS boundary in the context
of the AdS/CFT correspondence and holography.
E.g. three-dimensional AdS space  possesses
an infinite-dimensional, asymptotic  $W$-symmetry \cite{Brown:1986nw}, which is holographically realized as symmetry group of the two-dimensional conformal field theory that lives on the
boundary of the AdS space. In four space-time dimensions, the  BMS supertranslations were already introduced in 1962 by Bondi, 
van der Burg, Metzner and Sachs \cite{bms} (for more work on asymptotic symmetries in gravity see \cite{asym}).
These infinite-dimensional BMS transformations
describe the  symmetries of asymptotically-flat space-times at future or past 
null-infinity, denoted by ${\mathscr I}^+$  and ${\mathscr I}^-$ respectively. Hence the BMS groups 
$BMS^\pm$ describe the 
symmetries on ${\mathscr I}^+$  and ${\mathscr I}^-$, but in general not on the interior of four-dimensional space-time. Furthermore, if one considers a gravitational scattering process
(an S-matrix in Quantum
Field Theory) on asymptotically flat spaces, there is a non-trivial intertwining between $BMS^+$ and $BMS^-$ (see first reference in \cite{asym}).

However holography using black holes is independent of asymptotic boundaries but originates from considerations of the black hole horizon. 
This suggests that a more general set up should be
possible, extending the concept of symmetries also to the horizon of black hole.
Recently it was conjectured \cite{Strominger:2013jfa} (for other recent work on BMS symmetries, gravitational memory and the relation to soft theorems in scattering
amplitudes see \cite{strom})
that the BMS symmetry will play an important role in resolving the so called black hole information paradox by providing an (infinite-dimensional) hair, i.e.
charges 
to the black hole, that carry the information about the collapsing matter before the black hole is formed. Moreover it was argued \cite{Hawking:2015qqa} that the BMS group can can be extended
as symmetry group  $BMS^{\cal H}$ to the horizon ${\cal H}$ of a Schwarzschild black hole. 
However the relation between $BMS^{\cal H}$ and the standard BMS groups $BMS^\pm$ is still rather unclear. Also what is the action of $BMS^{\cal H}$ on the mass of the black hole
was so far not yet discussed.

In this paper we will explicitly construct the supertranslations $BMS^{\cal H}$ on the event horizon ${\cal H}$ of a Schwarzschild black hole with metric written in Eddington-Finkelstein coordinates\footnote{For additional recent work on infinite (BMS-like) symmetries on black hole horizon see \cite{bmshor}.} In addition there are still the  standard BMS transformation $BMS^-$ acting on the past null infinity surface ${\mathscr I}^-$ of the black hole geometry. 
Working out the relation between $BMS^{\cal H}$ and  $BMS^-$,
it is possible
to entangle  the transformations $BMS^{\cal H}$ and $BMS^-$ in  a non-trivial way. Namely we will show that the horizon supertranslations $BMS^{\cal H}$ 
contain a part that cannot be compensated by $BMS^-$ transformations. This part of supertranslations
on ${\cal H}$, which can be 
formally denoted by $\cal A \equiv BMS^{\cal H}/BMS^-$, act only on the horizon of the black hole and take the form of area-preserving diffeomorphisms on the black hole event horizon ${\cal H}$.
We will show that these $\cal A$-supertranslations leave the ADM mass of the black hole invariant. Using this information we can explicitly construct the Bogoliubov/Goldstone-type 
modes that correspond to the $BMS^{\cal H}/BMS^-$ transformations on the Schwarzschild metric, and show in this way that these modes are classically gapless.

As we shall argue, the transformations $BMS^{\cal H}/BMS^-$
are those that give raise to the microstates of the black hole that are relevant for the black entropy and eventually also for the solution of the black hole information puzzle.
Hence, the supertranslations $BMS^{\cal H}/BMS^-$ provide a gravitational hair on the black hole horizon.
Note that, as we will discuss, for infinite radius $r_S$ of the event horizon, the groups $BMS^{\cal H}$ and  $BMS^-$ become identical to each other; in this limit we are
dealing just with the BMS transformations of four-dimensional Minkowski space, since for $r_S\rightarrow\infty$ approaches the Minkowski metric (see the appendix of this paper).

  In the second part of the paper, we shall address the question of quantum information counting that is encoded in states created by the transformations  $BMS^{\cal H}/BMS^-$.  Classically, these transformations connect the different vacuum states of the black hole into each other and the corresponding classical Goldstone field is exactly gapless. Thus, classically the amount of entropy carried by these modes 
  is {\it infinite}. This nicely matches the fact originating from the microscopic theory \cite{Nportrait} that in classical theory 
  $N$ is infinite and so must be the amount of information carried by the black hole hair. 
 However, as we shall argue,  in quantum theory the Goldstone field 
 of $BMS^{\cal H}/BMS^-$-transformation
 is no longer gapless and contains only a finite number of Bogoliubov modes that can contribute into the entropy.   We shall show that the number of the eligible modes scales as $N$ and thus gives the correct scaling of black hole entropy.  
 
 We thus provide a crucial missing link between the geometric description of the pure gravitational hair of a black hole and its quantum entropy.   
 This result is in full agreement with the idea  of \cite{Dvali:2015rea} about the possible geometric interpretation of  Bogoliubov modes, but now we have an explicit candidate for it.  
 
   Finally, we devote a section in explaining the crucial role of quantum criticality in obtaining the correct entropy counting, once we go from classical to quantum regimes.  We explain, why naively dividing the horizon area into Planck-size pixels does not work for understanding the black hole entropy, and why one needs an input from 
   quantum criticality  \cite{gold,giamischa} for quantifying the amount of information carried by the black hole hair in the quantum world.

\section{Supertranslations on event horizon of a Schwarzschild-black hole}

\subsection{The standard BMS transformations on null infinity}

First let us review the classical BMS transformations \cite{bms,strom} in more detail.
As usual, in this context  we perform a coordinate transformation to Bondi coordinates by introducing a retarded or advanced time $u$ and spherical coordinates $r,\theta,\phi$:
\begin{eqnarray}
u&=&t\mp r\,, \nonumber\\
r\, \cos\theta&=&  x_1  \, , \nonumber\\
r\, \sin\theta e^{i\phi}&=& x_3+ix_2\, .
\end{eqnarray}
Instead of  $\theta$ and $\phi$ one can also use  complex coordinates $z,\bar z$ on the (conformal) sphere $S^2$:
\begin{equation}
z=\cot \biggl({\theta\over 2}\biggr)e^{i\phi}\, .
\end{equation}
In this coordinate system, future or past
null infinity, denoted by ${\mathscr I}^+$ or by ${\mathscr I}^-$ respectively, are the two null surfaces at spatial infinity:
\begin{equation} \label{skri}
{\mathscr I}^\pm\, :\quad R^1_{r=\infty, u}\otimes S^2_{\phi,\theta}\, .
\end{equation}
The future (past) boundaries of ${\mathscr I}^\pm$ ($r=\infty, u=\pm \infty, z,\bar z$) are denoted by ${\mathscr I}^\pm_\pm$.

 Asymptotically flat metrics in retarded (advanced)  time coordinates have an expansion around ${\mathscr I}^\pm$ with the following few terms:
\begin{eqnarray}\label{bondimetric}
ds^2&=&-du^2\mp dudr+2r^2\gamma_{z\bar z}dzd\bar z\nonumber\\
&~&{2m_B\over r}du^2+rC_{zz}^\pm dz^2+rC^\pm_{\bar z\bar z}d\bar z^2-2U^\pm_zdudz-2U^\pm_{\bar z}dud\bar z+\dots \, .
\end{eqnarray}
Here $\gamma_{z\bar z}$ is the round metric of the unit $S^2$ and
\begin{equation}
U^\pm_z=-{1\over 2}D^zC^+_{zz}\, .
\end{equation}
Furthermore, $m_B$ is the Bondi mass for gravitational radiation, the $C_{zz}^\pm$ are in general functions of $z,\bar z,u$  and the Bondi news $N_{zz}^\pm$ are characterizing outgoing (ingoing) gravitational waves:
\begin{equation}
N^+_{zz}=\partial_uC^+_{zz}\, .
\end{equation}
Gravitational vacua with zero radiation have $N^\pm_{zz}=0$.

$BMS^\pm$ transformations are defined as the subgroup of diffeomorphisms that acts non-trivially on the metric around ${\mathscr I}^\pm$, but still
preserve the asymptotic structure of the metric defined in eq.(\ref{bondimetric}).
They include Lorentz transformations and
an infinite family of supertranslations, which are generated by an infinite number of functions $g^\pm(z,\bar z)$.
Specifically the $BMS^\pm$ transformations act on the Bondi coordinates in the following way:
\begin{eqnarray}\label{bmstrans}
u&\rightarrow& u-g^\pm(z,\bar z)\, ,\nonumber\\
z&\rightarrow &z+{1\over r}\gamma^{z\bar z}\partial_{\bar z}g^\pm(z,\bar z)\, ,\nonumber\\
\bar z&\rightarrow &\bar z+{1\over r}\gamma^{z\bar z}\partial_{ z}g^\pm(z,\bar z)\, ,\nonumber\\
r&\rightarrow &r-D^zD_zg^\pm(z,\bar z)\, .
\end{eqnarray}
One can introduce  vector fields $\xi$, which generate infinitesimal $BMS^\pm$ supertranslations:
\begin{equation}
BMS^\pm \,:\quad \xi_g=g^\pm{\partial\over\partial u}+D^zD_zg^\pm{\partial\over \partial r}-{1\over r}(D^2g^\pm{\partial\over \partial z}+h.c.)\, .
\end{equation}

Then one can show that the supertranslations act on the Bondi mass and on 
$C_{zz}^\pm$ as
\begin{eqnarray}\label{supertrans}
{\cal L}_{\xi_g} m_B&=&g^\pm \partial_um_B\, ,\nonumber \\ 
{\cal L}_{\xi_g} C^+_{zz}&=&g^\pm\partial_u C^\pm_{zz}-2D^2_zg^\pm=g^\pm  N^\pm_{zz}   -2D^2_zg^\pm\, .
\end{eqnarray}

To each of these BMS transformation one can associate corresponding generators $T(g^\pm)$. These supertranslation generators act on the data around ${\mathscr I}^\pm$
as
\begin{eqnarray}\label{bmscharges}
\lbrace T(g^\pm),N^\pm_{zz}\rbrace& =&g^\pm \partial_uN^\pm_{zz}\, ,\nonumber\\
\lbrace T(g^\pm),C^\pm_{zz}\rbrace& =&g^\pm \partial_uC^\pm_{zz}   -2D^2_zg^\pm  \, .
\end{eqnarray}
Among themselves they build an infinite dimensional algebra, which is in fact closely related to the Virasoro algebra. 

It is important to emphasize that the BMS transformations are only asymptotic symmetries in the sense that the generators $T(g^\pm)$ correspond to global symmetries only on the
asymptotic null surfaces ${\mathscr I}^\pm$. In the interior of space-time the BMS transformations have to be viewed as local gauge transformations, i.e., here they just act as local
diffeomorphism on the space-time metric. Hence the globally-conserved BMS charges  only exist on ${\mathscr I}^\pm$ but not on the entire space-time.

Eq.(\ref{supertrans}) also means that super translations are spontaneously broken in the vacuum.
In field-theoretic language their action on the ground-state of the system leads to the generation of a Goldstone boson, namely the soft graviton.
One can also view this action as the transition between different BMS-vacua, which are separated from each other by a radiation pulse. The initial and final states precisely differ
by a BMS supertranslation, which is also known as the {\sl gravitational memory effect}.  This is also in agreement with the fact that the BMS transformations on ${\mathscr I}^\pm$ are global
symmetries of the field theory
S-matrix that describes the scattering of soft gravitons.

We can specialize the BMS transformations to Minkowski space-time.
In  terms of the Bondi coordinates, the Minkowski metric takes the following form:
\begin{eqnarray}\label{bondimetricMin}
ds^2=-du^2\mp dudr+2r^2\gamma_{z\bar z}dzd\bar z \, .
\end{eqnarray}

As seen from eq.(\ref{supertrans}), the super translations on Minkowski space do not generate a Bondi mass. However,
even starting with a trivial metric with $C^\pm_{zz}=0$, the supertranslations in general  generate a nontrivial $C^\pm_{zz}$:
\begin{equation}\label{ctrans}
C^\pm_{zz}\,\rightarrow \, C^\pm_{zz}-2D^2_zg^\pm\, .
\end{equation}
So the supertranslations do not act trivially on Minkowski space-time, but instead  generate an infinite 
family of Minkowski metrics which are all flat, i.e., all describe Minkowski space-time:
\begin{eqnarray}\label{bondimetricMina}
ds^2=-du^2-dudr+2r^2\gamma_{z\bar z}dzd\bar z
-2rD^2_zg^\pm dz^2-2rD^2_zg^\pm d\bar z^2 \, .
\end{eqnarray}
The BMS charges $T(g^\pm)$ are again localized only on ${\mathscr I}^\pm$ of Minkowski space, and not in its interior.

\subsection{The supertranslations on the black hole horizon}

A review on the geometry and physics of null infinity can be found in \cite{Ashtekar:2014zsa}. There, the BMS-group is defined intrinsically on (future or past) null infinity in a coordinate-independent manner. Since the null infinity in a conformal completion shares many similarities with a black hole event horizon, it is tempting to extend this definition to the case of an event horizon. Indeed, in \cite{Hawking:2015qqa}, there was made the proposal that the group of supertranslations on the event horizon plays an important role in the resolution of the black hole information paradox. In this note, we wish to identify the right supertranslation group, responsible 
for black hole information and show that the corresponding Goldstone modes are indeed gapless and have the right properties to match the critical Bogoliubov
modes of quantum theory.   

 We show that existence of these classically-gapless modes is intrinsically linked with  the existence of the event horizon (and the associated supertranslation group). 
  We hence establish the mapping between the existence of the horizon in the classical description and an underlying quantum critical state, as it was explained in the introduction.    
 
 Although the reasoning is extendable to general horizons, for concreteness and simplicity, we shall consider the case of an eternal Schwarzschild black hole. 

Consider the Schwarzschild-metric in (infalling) Eddington-Finkelstein coordinates $(v,r,\theta,\phi)$ given by the coordinate-transformation
\begin{equation}
v= t + r^* 
\end{equation}
with 
\begin{equation}
dr^* = (1-\frac{r_S}{r})^{-1} dr \,.
\end{equation}
This choice of Eddington-Finkelstein coordinates covers the exterior and the future-interior of a Schwarzschild black hole and the metric gets the form
\begin{equation}
ds^2 = -(1-\frac{r_S}{r})dv^2 + 2dvdr + r^2 d\Omega^2\,.
\end{equation} 

On the event horizon $\cal H$ with  $r=r_S$  we have a nullvector $n^\mu=(1,0,0,0)$ with $g_{\mu \nu}n^{\mu}n^{\nu}=0.$ We define the group of supertranslations on the horizon ${\cal H}$ by its Lie-Algebra  in analogy with the way the group of supertranslations is defined on null-infinity (see the definition in \cite{Ashtekar:2014zsa}). This Lie-algebra consists of vector fields $\zeta^\mu$ on the entire space-time, which on the horizon are supposed to satisfy the condition $\zeta^\mu |_{r=r_S} = fn^\mu$, where $f$ is a real function on the horizon with the Lie-derivative in null-direction vanishing, i.e.,  $L_n f=0$. The vector fields with $\zeta^\mu |_{r=r_S} = 0$ are divided out. 
In Eddington-Finkelstein coordinates representatives in the Lie algebra have therefore the form 
\begin{equation}
\zeta^\mu = (f,A,B,C)\, ,
\end{equation}
with the components satisfying the boundary conditions
\begin{align}
f|_{r=r_S} = f(\theta, \phi)\, , \label{RB}\\
A|_{r=r_S} = B|_{r=r_S} = C|_{r=r_S} = 0.
\label{Randbedingungen}
\end{align}
We now choose the representatives of the equivalence classes in the Lie algebra by specifying four coordinate conditions for the metrics we will consider in a moment
\begin{align}
g_{10} &= 1\, ,\\
g_{1i} &= 0 \,.
\end{align}

Remember that in a Hamiltonian formulation of the Einstein-Hilbert action, the coordinate conditions have to be imposed to fix the gauge-freedom. The coordinate conditions help in eliminating redundant canonical coordinates/momenta in order to leave a minimal set of two pairs of canonically-conjugated coordinates and momenta (two helicities of the graviton). For a review of the Hamiltonian formulation of general relativity, see \cite{Arnowitt:1962hi}. 

By choosing these coordinate conditions we specify the coordinate-system under the consideration. We require that coordinate transformations induced by the concrete representatives 
$\zeta$ do not leave this particular choice of coordinates, i.e., we require
\begin{equation}
\delta_{\zeta} g_{1 \mu} = 0.
\label{Bestimmungsgleichung}
\end{equation}

This requirement ensures that the induced field excitation $\delta_{\zeta} g_{\mu \nu}$ is physical. It is a shift in the phase space spanned by the minimal set of four canonically conjugated variables. For an observer in this coordinate conditions, 
$\delta_{\zeta} g_{\mu \nu}$ is a measurable excitation of the metric field rather than a redundancy. This observation is crucial in the argument showing that black holes do carry classical hair. Furthermore, precisely this requirement fixes the representative of each equivalence class in the Lie algebra.
 Eq.\eqref{Bestimmungsgleichung} yields the first order differential equations for the components 
\begin{align}
\frac{\partial f}{\partial r} &= 0\, ,\\
\frac{\partial A}{\partial r} &= 0\, ,\\
\frac{\partial}{\partial r}(r^2 B) + \frac{\partial f}{\partial \theta} - 2rB &=0\, ,\\
\frac{\partial}{\partial r}(r^2 \sin^2 (\theta) C) + \frac{\partial f}{\partial \phi} - 2r\sin^2 (\theta) C &= 0.
\end{align}
Together with the boundary conditions \eqref{RB}, \eqref{Randbedingungen} this fixes the representatives uniquely, and the supertranslations $BMS^{\cal H}$ on the horizon ${\cal H}$
are generated by the following vector fields $\zeta^\mu$, which can be labeled by function $f=f(\theta, \phi)$ on a two-sphere $S^2$:
\begin{equation}
BMS^{\cal H}\, :\quad \zeta_f^\mu = \biggl(f(\theta,\phi), 0, \frac{\partial f}{\partial \theta}(\frac{1}{r}-\frac{1}{r_S}), \frac{1}{\sin^2 \theta} \frac{\partial f}{\partial \phi}(\frac{1}{r}-\frac{1}{r_S})\biggr).
\label{Supertranslationen}
\end{equation}

\subsection{BMS-supertranslations of the Schwarzschild metric on   ${\mathscr I}^-$ }

Taking the limit $r_S \to \infty$ takes the future horizon $r=r_S$ to past null infinity $\mathscr{I}^-.$ Therefore, taking this limit in \eqref{Supertranslationen} 
yields the BMS-supertranslations at null infinity $\mathscr{I}^-$ with respect to the particular coordinate conditions used in the previous section:
\begin{equation}
\eta_g^\mu = \biggl(g(\theta,\phi), 0, \frac{\partial g}{\partial \theta}\frac{1}{r}, \frac{1}{\sin^2 \theta} \frac{\partial g}{\partial \phi}\frac{1}{r}\biggr)
\label{BMS}
\end{equation}
Therefore, in addition
 to the supertranslations eq.\eqref{Supertranslationen} on the event horizon ${\cal H}$ labeled by $f=f(\theta, \phi),$ there exist (for all isolated systems) the standard 
$BMS^-$-supertranslations at null infinity \eqref{BMS} labeled by an independent function $g=g(\theta, \phi)$.

 Let us remark that taking the limit $r_S \to \infty$ we are not led to consider the standard BMS transformations on ${\mathscr I}^+$. 
So for the black hole metric, we are still considering two {\it a priori} independent supertranslations, namely,
$BMS^{\cal H}$ on the horizon and $BMS^-$ on past null infinity ${\mathscr I}^-$.
Hence instead of the standard supertranslations $BMS^+$ we are dealing in case of the black hole with  $BMS^{\cal H}$, whereas the group $BMS^-$ still provides
the asymptotic symmetries of the black hole at ${\mathscr I}^-$.
 In other words, the dynamics of black hole formation is  associated with the relation between the two BMS groups we are dealing with, namely $BMS^-$, describing the collapsing body before black hole formation,  and $BMS^{\cal H}$ associated with the horizon of the black hole.

\section{Bogoliubov-modes as classical black hole hair}

Associated to the supertranslations \eqref{Supertranslationen} we can calculate the Bogoliubov-modes $\delta_\zeta g_{\mu \nu}$. 
Note that any representative of an equivalence class in the Lie algebra of supertranslations, i.e., any vector field satisfying the boundary conditions \eqref{RB}, \eqref{Randbedingungen}, would give rise to solutions $\delta_\zeta g_{\mu \nu}$ of the Bogoliubov-de Gennes equations. However, for an observer in a concrete coordinate system we have to specify the additional coordinate conditions (i.e. fix the gauge-freedom). Bogoliubov modes which do not fulfill the coordinate conditions (called ghosts) are not the observable excitations for an observer in this particular coordinate system. In that particular gauge, ghost do not correspond to shifts in the phase space spanned by the gauge-fixed minimal set of canonical variables and are therefore unphysical. The supertranslations derived by us (both on null infinity as well as on the event horizon), by-construction, induce physical excitations (see also the discussion in the last section). For our coordinate conditions the supertranslations $BMS^{\cal H}$ have the form \eqref{Supertranslationen} and the induced physical Bogoliubov-excitations have the form

\begin{align}
\delta_{\zeta_f} g_{\mu \nu}
=
\begin{pmatrix}
0 & 0 & -(1-\frac{r_S}{r})\frac{\partial f}{\partial \theta} & -(1-\frac{r_S}{r})\frac{\partial f}{\partial \phi}\\
0 & 0 & 0 & 0\\
* & * & 2r^2(\frac{1}{r}-\frac{1}{r_S})\frac{\partial^2 f}{\partial \theta^2} & 2r^2(\frac{1}{r}-\frac{1}{r_S})(\frac{\partial^2 f}{\partial \theta \partial \phi}-\cot \theta \frac{\partial f}{\partial \phi})\\
* & * & * & 2r^2(\frac{1}{r}-\frac{1}{r_S})(\frac{\partial^2 f}{\partial  \phi^2} + \sin \theta \cos \theta \frac{\partial f}{\partial \theta})
\end{pmatrix}
.
\label{Bogoliubov}
\end{align}  

On the other hand, the Bogoliubov excitations with respect to the standard $BMS^-$-supertranslations in eq.\eqref{BMS} of the Schwarzschild-metric in Eddington-Finkelstein coordinates are given by

\begin{align}
\delta_{\eta_g} g_{\mu \nu}=
\begin{pmatrix}
0 & 0 & -(1-\frac{r_S}{r})\frac{\partial g}{\partial \theta} & -(1-\frac{r_S}{r})\frac{\partial g}{\partial \phi}\\
0 & 0 & 0 & 0\\
* & * & 2r \frac{\partial^2 g}{\partial \theta^2} & 2r (\frac{\partial^2 g}{\partial \theta \partial \phi}-\cot \theta \frac{\partial g}{\partial \phi})\\
* & * & * & 2r (\frac{\partial^2 g}{\partial  \phi^2} + \sin \theta \cos \theta \frac{\partial g}{\partial \theta})
\end{pmatrix}\, .
\label{BMS-Bogoliubov}
\end{align}

\subsection{Microstates of Schwarzschild black hole}

Comparing the Bogoliubov-excitations due to supertranslations on the event horizon \eqref{Supertranslationen} with the excitations due to ordinary BMS-supertranslations \eqref{BMS}, we see that the horizon supertranslations contain a part which cannot be compensated by standard BMS-supertranslations. The horizon supertranslations therefore provide new Bogoliubov-excitations which are intrinsically due to the presence of an event horizon. In order to factor-out the part of event horizon supertranslations, which is not due to 
standard BMS-supertranslations, we define "`disentangled"' supertranslations with respect to the  quotient space ${\cal A} \equiv BMS^{\cal H}/BMS^-$. These are defined by 
the vector field which can be formally seen as the difference of the two vector fields, $\zeta$ and $\eta$, where we set the functions $g$ and $f$ equal each to each other.
In this way we obtain the vector field $\chi$ as
\begin{eqnarray}\label{quotient}
BMS^{\cal H}/BMS^-\, :\quad \chi_f^\mu& =& \zeta_f^\mu - \eta_{f}^\mu\nonumber\\
&=&(0, 0,-\frac{1}{r_S} \frac{\partial f}{\partial \theta}, -\frac{1}{r_S\sin^2 \theta} \frac{\partial f}{\partial \phi}) \,,
\end{eqnarray} 
for an arbitrary real function $f=f(\theta, \phi)$.
The corresponding Bogoliubov excitations are
\begin{align}
\delta_{\chi_f}g_{\mu \nu}=
\begin{pmatrix}
0 & 0 & 0 & 0\\
0 & 0 & 0 & 0\\
* & * & -2r^2 \frac{1}{r_S} \frac{\partial^2 f}{\partial \theta^2} & -2r^2 \frac{1}{r_S} (\frac{\partial^2 f}{\partial \theta \partial \phi}-\cot \theta \frac{\partial f}{\partial \phi})\\
* & * & * & -2r^2 \frac{1}{r_S} (\frac{\partial^2 f}{\partial  \phi^2} + \sin \theta \cos \theta \frac{\partial f}{\partial \theta}).
\end{pmatrix}
\label{microstates}
\end{align}  
Note that in the limit $r_S \to \infty$ the vector fields $\chi^\mu$ tend to zero and the horizon supertranslations become indistinguishable from the ordinary BMS-supertranslations. We shall call these additional Bogoliubov-modes \eqref{microstates} and \eqref{quotient}, which are due to the group $\cal A$, as $\cal A$-modes. Since the BMS-Bogoliubov modes, i.e., the modes due to BMS-supertranslations at null infinity $\mathscr I^-,$ are physical, they provide, already at the classical level, a hair to the black hole. In order to determine the point in phase space corresponding to the black hole state, the function $g$ in \eqref{BMS-Bogoliubov} has to be specified. However, this BMS-degeneracy is present for all isolated systems. Therefore, it is not to expect that these BMS-Bogoliubov modes contribute to the Bekenstein-Hawking entropy. However, there are additional physical excitations due to the $\cal A$-modes. It is highly expected that they contribute to the Bekenstein-Hawking entropy and should be identified with the black hole micro-states, as the presence of $\cal A$ is intrinsically due to the presence of the event horizon. In the next section, we will show explicitly that the $\cal A$-modes are classically-gapless, i.e., they do not change the value of the ADM-mass.\footnote{That the BMS-Bogoliubov-modes are gapless is well-known and follows from the fact that \eqref{BMS-Bogoliubov} \newline
decays as $O(\frac{1}{r})$ in cartesian-coordinates \cite{Arnowitt:1962hi}.} The group $\cal A$ provides additional hair to the back hole. One has to specify (independently of $g$) an additional function $f$ in \eqref{microstates} in order to determine the exact position of the black hole state in phase space. Therefore, we have two degeneracies providing classical black hole hair, which consistently collapse to ordinary BMS-degeneracy in the limit $r_S \to \infty$ in which the group $\cal A$ disappears.

\subsection{Gaplessness of $\cal A$-modes - Quantum Criticality}

In order to see that the Bogoliubov-excitations \eqref{microstates} are gapless, we translate them via $t=v-r^*$ to Schwarzschild coordinates $(t,r,\theta,\phi)$, which yields
\begin{eqnarray}
g_{\mu \nu} + \delta_{\chi_f} g_{\mu \nu}
={\tiny \begin{pmatrix}
-(1-\frac{r_S}{r}) & 0 & 0 & 0\\
0 & (1-\frac{r_S}{r})^{-1} & 0 & 0\\
0 & 0 & r^2 -2r^2 \frac{1}{r_S} \frac{\partial^2 f}{\partial \theta^2} & -2r^2 \frac{1}{r_S} (\frac{\partial^2 f}{\partial \theta \partial \phi}-\cot \theta \frac{\partial f}{\partial \phi})\\
0 & 0 & * & r^2 \sin^2 \theta -2r^2 \frac{1}{r_S} (\frac{\partial^2 f}{\partial  \phi^2} + \sin \theta \cos \theta \frac{\partial f}{\partial \theta})
\end{pmatrix}}
.
\label{microstates2}
\end{eqnarray}  

Next we compute the ADM-mass (in units where $G_N=1$) via a version of the Brown-York formula due to Brewin \cite{Brewin:2006qe}
 \begin{equation}
M_{ADM} = \frac{1}{8\pi} \lim_{S \to i^0} (\frac{dA}{ds})_+ - (\frac{dA}{ds})_-,
\end{equation}
where $S$ is the sphere $t=const., r=const.$ and $r$ is taken to infinity. In the expression above, outside the sphere (+) the space is taken to be flat and inside (-) the space is given by the actual metric. $A$ refers to the area of the sphere and $s$ to the proper length as measured with the line element. The induced metric on $S(r)$ is given by
\begin{align}
s_{ij}=
\begin{pmatrix}
r^2 -2r^2 \frac{1}{r_S} \frac{\partial^2 f}{\partial \theta^2} & -2r^2 \frac{1}{r_S} (\frac{\partial^2 f}{\partial \theta \partial \phi}-\cot \theta \frac{\partial f}{\partial \phi})\\
* & r^2 \sin^2 \theta -2r^2 \frac{1}{r_S} (\frac{\partial^2 f}{\partial  \phi^2} + \sin \theta \cos \theta \frac{\partial f}{\partial \theta})
\end{pmatrix}
.
\end{align}  
 Already from the form of $\chi^\mu$ it is clear that one can do a reparametrization in $(\theta, \phi)$ such that one obtains the line element of an Euclidean $S^2.$ 
 One has therefore
\begin{equation}
A_+(r) = A_-(r) = 4\pi r^2.
\end{equation}

Futhermore, one has
\begin{align}
ds_+ &= dr\, ,\\
ds_- &= (1-\frac{r_S}{r})^{-\frac{1}{2}}dr = (1+\frac{r_S}{2r})dr + O(\frac{1}{r}).
\end{align}
Altogether this yields
\begin{equation}
M_{ADM} = \frac{r_S}{2} = M.
\end{equation}

Therefore, for every Bogoliubov-excitation \eqref{microstates} given by a function $f=f(\theta, \phi)$ the ADM-mass is equal to the Schwarzschild-mass and therefore these excitations are gapless. It might look strange that the Bogoliubov excitations look as angular reparametrizations. However, the important point is that precisely those angular reparametrizations which are of the form given by $\chi^\mu$ are the physical Bogoliubov excitations. Although,  arbitrary angular reparametrizations solve the Bogoliubov-de Gennes equations, they are in general unobservable. Indeed, they also solve the coordinate condition we have imposed. But, since in our coordinate conditions we have fixed a time-component of the metric, one of our coordinate conditions fixes only a Lagrange-parameter of the ADM-formulation of general relativity. This Lagrange-parameter restricts one of the other parameters, but does not help in getting rid of one degree of freedom. However, the argument can be repaired by using coordinate conditions which fix four degrees of freedom, in such a way  that the remaining gauge freedom is really due to physical excitation of the canonical variables. One then carries out the same argument (with more involved  equations), transforms back to our coordinate conditions and arrives  to the same result that precisely the Bogoliubov excitations given above correspond to the physical gapless excitations. The presence of the physical gapless angular reparametrizations is due to the presence of the black hole event horizon. 

Thus, we have shown that the $\cal A$-modes are gapless. The presence of these additional gapless excitations is a highly non-trivial phenomenon and 
is the classical sign of  the underlying quantum criticality of the Schwarzschild geometry. 
 This phenomenon has several important consequences. In particular 
 it is the key to the information storage capacity of the black hole and respectively   
to the existence of black hole hair and entropy.  We shall quantify some of  these properties in the next section.

\section{Information Resolution and Entropy Counting} 

  We have identified the black hole hair carried by the Bogoliubov excitation \eqref{microstates} given by a function $f=f(\theta, \phi)$. We have shown that for this 
 form of excitations the ADM-mass is equal to the Schwarzschild-mass and therefore these excitations are classically-gapless.
 
  We now need to understand how much information can be encoded in these excitations.  Notice, that up until now, our analysis was purely classical. 
  That is, we have been working with finite $M$ and $G_N$, but zero 
  $\hbar$.  This is equivalent to saying that we were working with 
  $L_P = 0$ and thus, $N = \infty$.   The results obtained in this limit pass the  first consistency check that we are on the right track of understanding the origin of black hole entropy, 
 since as we saw, in this limit an {\it arbitrary} 
  deformation of the form $f=f(\theta, \phi)$ is exactly gapless.  Thus, 
  we can store an {\it infinite}  amount of information in such deformations. 
  We have thus correctly identified the source of infinite entropy in the limit 
  of $\hbar = 0$.

    We now need to understand the effects of finite $L_P$ (i.e., finite $N$). 
  For this we need to clarify the meaning that the  transformation 
  $f=f(\theta, \phi)$ acquires in the quantum theory.  For this let us perform 
  expansion of this function in spherical harmonics, 
  \begin{equation}
 f(\theta, \phi) = \sum_{l=0}^{l=\infty} \sum_{m=-l}^{m=+l} \, b_{lm} Y_{lm}(\theta, \phi), 
\end{equation}
where $b_{ml}$ are expansion coefficients.  Classically, these coefficients are 
just $c$-numbers, but in quantum theory they acquire  
 the meaning of (expectation values of) the occupation numbers of spherical momentum modes, $\hat{b}_{lm}$. 
 Notice that, since the only scale entering in (\ref{quotient}) and (\ref{microstates}) is 
 $r_S$, it is natural to think of the modes $b_{lm}$ as living on a sphere of radius 
 $r_S$ and correspondingly to measure the momenta in units of ${\hbar \over r_S}$.
 
  Thus, by promoting the expansion coefficients $b_{lm}$ into the expectation values of operators, we establish connection between the classical and quantum pictures.  
 In classical theory, by performing $BMS^H/BMS^-$ transformation, parameterized by the function $f=f(\theta, \phi)$, we excite the classical  Goldstone field.  In quantum theory, the same transformation is equivalent of 
 populating system with different occupation numbers of $\hat{b}_{ml}$-modes.

  An each momentum mode in quantum theory acts as a qubit that can store 
  information in its occupation number, $n_{lm} \equiv \langle 
 \hat{b}_{lm}^{+} \hat{b}_{lm} \rangle$.   These quantum modes are nothing but the Bogoliubov modes appearing at the quantum critical point.  The amount of the information carried by the black hole hair is determined by the information-storage capacity of these modes. In order to measure it we need to answer the following question:  \\

{\it How many independent angular Bogoliubov qubits must be counted as the legitimate information carriers? } \\ 
 
   Let  $l_{max}$ be a maximal spherical harmonic number up to which we include 
   modes in entropy-counting.  Then, the total number of included modes  scales as
   $\sim l_{max}^2$ and the major contribution into this number comes from the highest harmonics, i.e., modes with $l \sim l_{max}$.   The value of 
   $l_{max}$ can be estimated from the requirement that the energy gap, $\Delta E$, 
 in any given mode contributing to the entropy-counting must be bounded by, 
\begin{equation}
 \Delta E  \lesssim  {1\over N} {\hbar \over r_S}\,. 
 \label{boundE}
 \end{equation} 
   The energy gap generated by the quantum effects in a given mode of momentum 
 $l {\hbar \over r_S}$ is proportional to the strength of gravitational self-coupling $\alpha(l) \equiv  l^2 {L_P \over r_S^2}$ as well as to the departure of the background from criticality, 
i.e., $(1-\alpha(r_S) N)$.  Exactly at criticality we have
$\alpha(r_S) = {1\over N}$. However, the quantum fluctuations in occupation number 
cause small departure from the critical relation. The effect can be estimated as 
$(1-\alpha(r_S) N) \sim {1\over N}$. 
We thus arrive to the following estimate, 
$\Delta E \sim {\hbar \over r_S} \alpha(l) (1 - \alpha(r_S) N) \sim  {l^2\over N^2} {\hbar \over r_S}$.   Inserting this into (\ref{boundE}) we get the estimate 
  $l_{max} \lesssim \sqrt{N}$.     
 Thus, the total number of Bogoliubov qubits contributing into the entropy scales as 
 $\sim l_{max}^2 \sim N$.   
  As discussed in the 
 introduction, this is exactly the number of information carriers needed for accounting  for Bekenstein entropy! 
  
  We thus conclude, that in quantum theory, the amount of information carried by the black hole hair generated by $BMS^{\cal H}/BMS^-$ transformations is finite 
  and is given by $N$. 
  
  We would like to stress the striking similarity between the counting obtained above 
  and the similar counting obtained using the so-called St\"uckelberg approach  for accounting the quantum information in the presence of a horizon \cite{stuckelberg}. 
  This is probably not surprising, since the vector field $\chi_{\mu}$ and the corresponding Goldstone field $f(\theta,\phi)$ can be thought to be related with the 
  St\"uckelberg field, discussed in \cite{stuckelberg}, needed for maintaining the gauge invariance from the point of view of the external observer in the presence of the information boundary.  The role of the latter in the present case is played by the 
  two-sphere that is subjected to  $\cal A$-transformations.

\section{Crucial Role of Quantum Criticality in Entropy Scaling} 

 Although this is already clear from our previous analysis,  we would like to explain
why quantum criticality is absolutely crucial for obtaining the correct scaling of entropy,  when we go from the classical ($\hbar =0,~ L_P=0$) to the quantum 
($\hbar \neq 0,~ L_P\neq 0$) 
description. 
 In particular, we would like to explain why explaining entropy by a  simple division of the system into the Planck size pixels does not work.  
 
 As we have seen, in the classical theory the gapless excitation is given by a Goldstone field $f(\theta,\phi)$. Since $f$ is an arbitrary continuous function of spherical coordinates, we can encode an { \it infinite}  amount of information in it at {\it zero}  energy cost.  Indeed, in classical theory we can encode a message by giving different values to $f$ for set of coordinates, say, $\theta_1,\phi_1$ and 
 $\theta_2,\phi_2$ that can be {\it arbitrarily} close. This is because in classical theory the
 arbitrarily small coordinate differences  (i.e., $ \Delta \theta = \theta_2 - \theta-1, ~~
 \Delta\phi = \phi_2-\phi_1$) are resolvable at zero energy cost.  
 This is why the ground-state of any classical system with a  gapless field $f(\theta,\phi)$  has infinite entropy, from the information point of view. 

 In quantum theory, there is no longer a free lunch, because the coordinate-resolution costs finite energy. The amount of entropy carried by the same system, once $\hbar$ is set to be non-zero,  crucially depends on the energy-cost of the coordinate-resolution.   
This is why, {\it a priory} it is not obvious what shall remain out of the infinite classical entropy in the quantum world. 

   Without having a microscopic quantum theory that could tells us what is the
   energy-cost of coordinate-resolution, we can only make some guesses.  
For example, we can take a naive root and put a Planck-scale resolution cutoff on the 
coordinate-separation, demanding $r_S\Delta \theta > L_P, ~~r_S\Delta \phi > L_P$. 
That is, we decide to use in information-count only the values of the function $f$ in the set    
of points that are separated by distance of Planck length or larger.  This way of entropy-counting is equivalent 
to dividing the are of the sphere in Planck-area pixels, each housing a single quantum degree of freedom (i.e., a qubit). Since the total number of pixels is 
$N = R^2/L_P^2$, it is obvious that the scaling of 
entropy obtained in this way shall match the Bekenstein entropy.   
 This is essentially the approach originally introduced by 't Hooft \cite{thooft} and 
 also adopted recently  in \cite{Hawking:2016msc}.   
 
  The problem, however, is that such a method, despite being an interesting
 prescription,  cannot be considered, without an additional microscopic input,  
  to be an explanation of entropy scaling, because of the reason that we shall now explain. 
 
  In a generic quantum system, housing a degree of freedom (say a two-level quantum system) in a Planck-size pixel would cost the energy gap 
  $\Delta E \sim {\hbar \over L_P}$. Since the total number of pixels, is $N$,  
 the resulting number of states, $2^N$, would naively account for Bekenstein entropy.  However,  these states  span an enormous total energy gap 
 \begin{equation}
 \Delta E_{total} \sim  N {\hbar \over L_P}\, ,
 \label{totalGap}
 \end{equation}
  which is by a factor $\sqrt{N}$ larger 
 than the black hole mass!
  This, of course, shows that  such states cannot exist and cannot be counted in entropy.  
 
  So, in order to account for the entropy-counting, we must explain how come the 
  energy gap per pixel, instead of being  $\Delta E \sim {\hbar \over L_P}$, is  
  \begin{equation}
  \Delta E \sim {1 \over N^{{3\over 2}}}{\hbar \over L_P}\,.
  \label{truegap}
  \end{equation} 
   The additional suppression factor $N^{-{3\over 2}}$ is enormous even for 
   relatively-small black holes. For example, for a black hole of earth's mass this factor would be $N^{-{3\over 2}}\sim 10^{-99}$. 
   
     This enormous suppression is precisely what quantum criticality delivers, as 
  explained in series of papers \cite{gold,giamischa} and as outlined in the previous section. 
  For a detailed review of the role of quantum criticality for obtaining such cheap qubits, we refer the reader to \cite{giamischa} and references therein.

\section{Gravitational vacua, graviton condensates and Goldstone modes}

\subsection{Minkowski vacua}

 The action of  BMS transformations on some gravitational metric $G$ reveals  the existence of  different gravitational vacua we can denote by  $|g;G\rangle$ for $g$ being the BMS function.
The BMS transformations $\xi_g$ normally act on the null surfaces of asymptotically flat metrics at spacial infinity.
This applies also to Minkowski space time, and hence there exist {\it infinitely many} distinct Minkowski vacua, $|g^-;Minkowski\rangle$,
which correspond to the family of BMS-transformed metrics as given by equation (\ref{bondimetricMina}).

In \cite{Dvali:2015rea} we have explained that this degeneracy of Minkowski vacua  
can be understood by describing Minkowski as a coherent state of $N=\infty$ gravitons of {\it infinite} wavelength. 
This  behavior naturally emerges in the  framework 
of  the black hole quantum N-portrait \cite{Nportrait}, in which a black hole is described as a state of  $N$ soft gravitons (coherent state or a Bose-Einstein condensate) at a quantum critical point, $\alpha N = 1$.  

 
On the other hand, one can show (see the appendix) that the {\it near horizon geometry} in the infinite mass limit of a Schwarzschild black hole is a Minkowski space. The value of this comment lies in relating the infinite number of Minkowski vacua we can define on null infinity with the limit of the black hole entropy in the limit of infinite mass. This is the limit in which the moduli of Minkowski vacua generated by asymptotic symmetries can match the expected entropy of an infinitely massive black hole. Namely, in this limit the horizon becomes the null infinity and both asymptotic symmetries $BMS^-$ and $BMS^{\cal H}$ become isomorphic.

As also discussed in  \cite{Dvali:2015rea} the supertranslations are spontaneously broken in the gravitational vacuum of the theory, in particular by the vacuum state that corresponds
to Minkowski space. So acting with the $BMS^-$ supertranslations on infinite-$N$ graviton state, that corresponds to Minkowski space,
the Goldstone equation has the following form:
 \begin{equation}\label{MinBMS}
BMS^-\, :\qquad  T(g^-)|Minkowski\rangle =|g^-;Minkowski\rangle\, .
\end{equation}
The corresponding Goldstone modes precisely correspond to the 
Bogoliubov excitations  eq.\eqref{BMS-Bogoliubov} with $r_S\rightarrow\infty$
that are given with respect to the standard $BMS^-$ transformations
 on ${\mathscr I}^-$.
 Equation   \eqref{MinBMS}  means that Minkowski space is infinitely degenerate and is described 
 by the infinite family of Minkowski metrics as given in eq.\eqref{bondimetricMina}.
The states $ |Minkowski\rangle$ and $|{g^-};Minkowski\rangle$ correspond to two Minkowski metrics that are related by a $BMS^-$ transformation and the difference among the two metrics
can be accounted by having two different functions $C^\pm_{zz}$, which are related by the $BMS^-$ transformation on $C^\pm_{zz}$, as shown in eq.\eqref{ctrans}.
 
  Note that what we learn from the classical analysis of the asymptotic symmetries is the existence of many classical vacua interpreted as connections defined by $C$ functions that are total derivatives $D_z^2g$.  In quantum theory the manifold of classically-inequivalent vacua corresponds to different quantum vacuum states. Each of these states is characterized by a BMS function $g$. Quantum mechanically  these vacua for different $g$'s are orthogonal. Moreover, action by a  broken symmetry transformation that connects two vacua is equivalent to  
creation of soft gravitons that play the role of the Goldstone bosons for the large diffs defining the asymptotic symmetry. When we talk about a quantum resolution of these vacuum states we really mean a coherent state representation of them in terms of quanta with infinite wavelength. In the case of Minkowski vacua this quantum  resolution because of infinite-$N$ and infinite wavelength is automatically consistent 
with quantum mechanical stability. This is not the case for the metrics 
which in quantum picture give finite $N$, such as the finite mass 
black holes.  Quantum mechanically, such states are not necessarily stable.

\subsection{Black hole vacua}

As we discussed, since the black hole horizon is also a null surface, we can extend the rules of asymptotic symmetries, vacua and Goldstone modes to this case too. The first important thing to be clarified is what is playing the role of manifold of {\it vacua}.  
Following our previous discussion, it is very natural to identify this manifold with the space of states generated by the action of $\cal A \equiv BMS^{\cal H}/BMS^-$.

 In the quantum portrait the black hole state $|BH\rangle$ is  given in terms of 
the state of $N$ gravitons at criticality, $\alpha N = 1$. This criticality manifests in the appearance of classically-gapless Bogoliubov modes that we have identified with the gapless modes obtained by acting with $\cal A$. 

The degeneracy of  $|BH\rangle$ originates from the action of the supertranslations  $\cal A$ on the horizon ${\cal H}$ , where we have factored out the
standard BMS transformations on ${\mathscr I}^-$. These superstranslations, are defined  in eq.\eqref{quotient} and act as follows:
 \begin{equation}\label{MinBMS}
{\cal A} \, :\qquad  T(f)|BH\rangle =|f;BH\rangle\, .
\end{equation}
The corresponding Bogoliubov modes in \eqref{microstates} are shown to be 
classically-gapless and preserve the ADM mass of the black hole.
Hence, we propose to identify the family of black hole vacua $|f;BH\rangle$ with the family of $\cal A$-transformed black hole metrics, as given in eq.(\ref{microstates2}),
with the gapless Bogoliubov modes corresponding to the Goldstone field.

For infinite $N$, which corresponds to classical limit,  the Goldstone  field is exactly gapless. 
However, as discussed above, for finite $N$ the energy gap $\Delta E\sim {1 \over N}$ is generated. Moreover, the effective number of modes that counts in the entropy is 
finite and also set by $N$.  
 
Therefore we would like to stress that there are effects, captured by the quantum $N$-portrait,  that are expected not to be visible in the classical BMS picture and their account requires the finite-$N$ resolution.  
 In particular, we expect that because of the interaction among different modes, 
 at finite $N$ the different would-be vacua obtained by action of $\cal A$
 are no longer orthogonal.  This overlap is a quantum ${1\over N}$ effect and is linked with 
 black hole evaporation.

\subsection{Some remarks on the relation to gravitational scattering amplitudes}

It has been known from the early days of quantum field theory that in theories with long-range forces, such as QED or gravity, the asymptotic states needed to define an 
IR-finite 
S-matrix are not the ones diagonalizing the free part of the hamiltonian, but instead the dressed states with infinite number of soft quanta. Generically, these dressed states are coherent states. This phenomenon underlines the standard recipes, such as the Bloch Nordsieck theorem in QED, for  dealing with IR divergences. A nice way for understanding the role of asymptotic symmetries in Minkowski space-time is simply in terms of symmetries governing the kinematics of these asymptotic dressed states. This is in essence the meaning of the relation between soft quanta theorems and asymptotic symmetries. 

In the case of gravity we can consider an ultra-Planckian scattering with large center of mass energy $\sqrt{s}$ at small impact parameter. In such conditions we expect to form a black hole in the final state (ignoring the subsequent evaporation). Within the classicalization scheme described in \cite{Dvali:2010jz}, the final state of such scattering process can be formally represented as some quanta of matter (e.g., the ultra-Planckian scattering of two electrons ) gravitationally dressed by $N$ soft gravitons with $N\sim sL_P^2$, i.e.,
\begin{equation}
|f; coh(N)\rangle \, .
\end{equation}
In other words the final state is gravitationally-dressed by $N$ soft gravitons of typical wave length $r_S\sim \sqrt{s} L_P^2$. These $N$ gravitons are organized in the form of a coherent state that we have denoted $|f;coh(N)\rangle$,
 where  $f$ denotes the soft graviton part. The in-state picture was thought from the point of view of the standard kinematics, defining the S-matrix in an asymptotically flat space, and therefore, the relevant kinematic asymptotic symmetry was BMS($\pm$). However, if a black hole is formed, the gravitational dressing of the final state  should be modified in order to take into the account the degeneracy of final states 
 obtained by the action of $\cal A$.  
  That is, the S-matrix element must be schematically replaced by, 
\begin{equation}
\int_{\cal A} |\langle in | \hat{S} t|f; coh(N)\rangle|^2 \,, 
\end{equation}
where the integral it taken over the quotient space $\cal A \equiv BMS^{\cal H}/BMS^-$ and $t$ representing the action of this quotient space on the IR gravitationally dressed state. This integral formally compensates the suppression factor $e^{-N}$ of the gravitational $2\rightarrow N$ part of the ultra-Planckian scattering. 

This formal picture provides a nice understanding of black hole formation in ultra-Planckian scattering \cite{Dvali:2014ila}  in the following sense. First of all, the UV/IR non-Wilsonian structure of gravity leads us to a gravitational dressing by $N$ soft quanta. The suppression factor is however compensated by the difference between the asymptotic group $BMS^-$ and the  $BMS^{\cal H}$. From the microscopic point of view defined by the $N$-portrait this {\it enhancement} of asymptotic symmetries simply reflects the quantum criticality of the many body graviton system.

Of course, the previous discussion simply intends to grasp the underlying structure. The key point is how the integral over $\cal A$ precisely compensates the suppression factor $e^{-N}$, or in other words, how do we get the right black hole entropy.  This was explained above, by showing that at finite-$N$  the number of relevant Bogoliubov modes contributing into the entropy scales precisely as $N$. This fact shall ensure that 
for finite $N$ the integral over  $\cal A$ gets reduced to a discrete sum over 
$\sim 2^N$  final micro-states.  This provides the needed compensation for the suppression factor $e^{-N}$.

\section{Conclusions}

 The goal of this paper was to explicitly identify a candidate symmetry that in the geometric description can account for the existence of the degenerated black hole (micro)states, which give rise to  black hole entropy.  The information stored in these 
 states is a manifestation of the black hole hair.  From the microscopic quantum theory the existence of such a hair was concluded some time ago \cite{Nportrait, gold}, but 
 its precise geometric description has not been found so far. 
 
 In \cite{Dvali:2015rea}, using the view of Minkowski space as of infinite-mass limit of a black hole, this symmetry was identified with BMS symmetry of Minkowski space.  Minkowski space was described as a coherent state with infinite occupation number of infinite-wavelength gravitons. The symmetry that  connects the inequivalent Minkowski vacua changes the occupation number of these soft   
gravitons in the coherent state. These gravitons play the role of Goldstone bosons
that carry infinite entropy of Minkowski space.

   In this paper, we have suggested that the right candidate for such a  symmetry
  for a finite-size black hole is the transformation corresponding to the quotient  $\cal A \equiv BMS^{\cal H}/BMS^-$. 
 Geometrically, we have identified the supertranslations on the black hole horizon as being  area-preserving transformations on the two-sphere.  
   We have performed various consistency checks showing that this symmetry indeed passes all the tests for describing the geometric limit of an underlying quantum critical degeneracy of black hole micro-states. 
   
   In particular, we showed that the variations of the black hole metric, $\delta_{\chi} g_{\mu \nu}$, induced by $\cal A$, are classically gapless, as they do not change 
   the ADM mass of the metric.  Classically, these gapless excitations can be interpreted as the Goldstone field (parameterized by an arbitrary function $f(\theta,\phi)$) that connects the degenerate black hole ground-states. In the quantum language, these excitations correspond to the Bogoliubov excitations that give rise to the black hole micro-states. 
   In short, we have identified some clear hints for the quantum criticality of the Schwarzschild-geometry.
   
    In the classical limit, the hair and the entropy carried by these gapless modes is infinite and information cannot be resolved during any finite time. 
     Thus, for the information-recovery the quantum effects are crucial.  
       
  We have then showed, confirming the suggestion of the previous work \cite{Dvali:2015rea},  that in quantum theory, the gap is necessarily generated due to ${1 \over N}$-effects.  Simultaneously, the number of independent modes that contribute into the entropy counting becomes finite and scales as $N$.  This matches the black hole entropy scaling.    
   
We have argued that
in gravity boundary null-surfaces are equipped with data that qualify as classical ground-states. In quantum picture these ground-states can be described as  coherent states composed out of soft background modes and are related by the action of the asymptotic symmetry groups. 
Once the non-zero interaction among the background quanta is taken into the account, the degeneracy is expected to be lifted and the majority of the ground-states becomes unstable.  This will manifest in radiation coming from the null boundary in a perfectly unitary form.
  Of course, as argued in \cite{Dvali:2015rea}, this effect is absent for Minkowski space, since the constituent quanta have infinite wave-lengths, and correspondingly,  they have vanishing quantum gravitational coupling. \\

   Finally, let us make a comment that should help the reader to catch the essence of 
   the transformation $\cal A$ and the key difference of our result 
   from the recent analysis in \cite{Hawking:2016msc}.  
   For this note that the analog of $\cal A$ in case of $U(1)$-gauge theory would be trivial. In this case one can easily observe that any large gauge transformation on $\cal H$ can be compensated by the corresponding one on $\mathscr I^-$, leading to a trivial $\cal A$, i.e.,  equal to the identity. The obvious reason for this is that in the gauge case, contrary to what we have shown is happening in the case of gravity, the large gauge transformations have no explicit dependence on the geometrical data defining the horizon. In other words, in the gauge case the Goldstone bosons associated with large gauge transformations are always $S$-matrix soft photons and the only possibility to use them as some form of horizon hair is by implementing (or implanting in the sense of \cite{Hawking:2016msc}) soft photons on the horizon. This is due to the fact that for the case of electro-magnetic large gauge transformation $\cal A$ is trivial.  
   
For gravity and its BMS transformations the situation is dramatically different in two important senses. First, the quotient between the transformations on $\mathscr I^-$ and $\cal H$ is already non-trivial on the basis of the geometry and secondly this quotient defines physical transformations. Note again the important difference with the gauge case where, if we use two different  gauge shift functions  (denoted in \cite{Hawking:2016msc} by 
$\epsilon$) on $\cal H$ and $\mathscr I^-$, the difference becomes a {\it gauge redundancy}. 

As we have lengthly discussed along the paper, the non-triviality of $\cal A$ in the case of gravity is what allows us to identify the purely-gravitational candidates to account for the black hole entropy, or equivalently, the purely-gravitational hair. A hair that is soft, but relative not to a notion of $S$-matrix softness, but to the intrinsic notion of softness (existence of gapless modes) determined by the black hole criticality.  In this sense, the $\cal A$-Goldstones  (generated by the  vector $\chi_{\mu}$) can be viewed as the St\"uckelberg fields \cite{stuckelberg}  that manifest physicality of this transformation. 
Note that in the case of charged black holes the explicit dependence of $\cal A$ on the charge (or charges) of the black hole appears through the geometric dependence of $\cal A$ on the gravitational radius.

A potentially-important general lesson we can extract from our findings is that what characterizes the ability of a null-surface $\cal H$ to work as holographic screen, is the ``rank'' of the corresponding $\cal A(\cal H)$. 

\section*{Acknowledgements}
We thank Ioannis Bakas and Alexander Gu\ss mann for many discussions. We also thank Nico Wintergerst and Sebastian Zell
for discussions on related topics. 
	The work of G.D. was supported by Humboldt Foundation under Alexander von Humboldt Professorship,  by European Commission  under ERC Advanced Grant 339169 ``Selfcompletion'' and  by TRR 33 "The Dark
Universe".
The work of C.G. was supported in part by Humboldt Foundation and by Grants: FPA 2009-07908, CPAN (CSD2007-00042) and by the ERC Advanced Grant 339169 ``Selfcompletion'' . The work of D.L. was supported by  the ERC Advanced Grant 32004 ``Strings and Gravity'' and also by TRR 33.

\section*{Appendix: Minkowski space as the near horizon geometry of the Schwarzschild metric}

Let us  derive the near horizon limit of the Schwarzschild black hole.
 The well-known black hole metric has the form
 \begin{equation}
 ds^2=-(1-r_S/r)dt^2+(1-r_S/r)^{-1}dr^2+r^2d\Omega^2\,. 
 \end{equation}
 We now introduce the  coordinate $\epsilon=r-r_S$ and for small $\epsilon$ we obtain the metric in the near horizon limit 
  \begin{equation}
 ds^2=-{\epsilon\over r_S}dt^2+{r_S\over\epsilon}dr^2+r_S^2d\Omega^2\,. 
 \end{equation}
 Next we introduce two new coordinates,
 \begin{equation}
 \rho=\sqrt{r_S\epsilon}\, ,\qquad \omega={t\over r_S}\, ,
 \end{equation}
 and we take the large 
  $N$ limit 
  \begin{equation}\label{largeN}
N\rightarrow\infty\, ,\quad r_S\rightarrow\infty\, , \quad M\rightarrow\infty\,.
\end{equation}
 such that $\rho$ and $\omega$ are finite for small $\epsilon$ reps. for large $t$.
In this limit  the entropy as well as mass and horizon become infinitely large.
 
 Then the near horizon metric finally becomes:
  \begin{equation}
 ds^2=-\rho^2d\omega^2+d\rho^2+\sum_{i=2,3}dx^idx_i\,. 
 \end{equation}
 This is the metric of 
$M^{1,1} \times R^2$, where $M^{1,1}$ is the 2-dimensional Minkowski space in Rindler coordinates $\omega$ and $\rho$. These are related to the flat Euclidean coordinates
as
\begin{equation}
t=\rho\,\sinh \omega\, ,\qquad x_1=\rho\,\cosh \omega\, .
\end{equation}
In Rindler space the horizon  ${\cal H}$ of the black hole is at $\rho=0$, called the Rindler horizon, which corresponds to the two light cones, $x_1=\pm t$, of $M^{1,1}$:
\begin{equation} \label{horizon}
{\cal H}\, :\quad R^1_{\omega}\otimes R^2_{x_2,x_3}=R^1_{x_1=\pm t}\otimes R^2_{x_2,x_3}\, .
\end{equation}

In summary, in its Rindler version Minkowski space-time is just the near horizon geometry of a black hole with 
\begin{equation}
N=M=r_S=\infty\, .
\end{equation}
As discussed before, in this limit the black hole horizon  ${\cal H}$ agrees with ${\mathscr I}^-$.
 Note
that the Rindler space with its horizon ${\cal H}$ radiates, it is a thermal space with some Rindler temperature. However, the 
Rindler entropy is infinite, just as the entropy of a infinitely-massive black hole.

In summary, we are obtaining here the picture, which is in close analogy with the emergence of $AdS$-geometry via the  branes in superstring or M-theory:
four-dimensional Minkowkski space $M^{1,3}$, written as $M^{1,1} \times R^2$,
arises as the near horizon geometry of a Schwarzschild black hole. In analogy to the $N$ coincident D3-branes of the $AdS_5\times S^5$ geometry, the microscopic picture of 
$M^{1,1} \times R^2$
 is the bound state (sort of a Bose-Einstein condensate) of $N$ graviton particles.
   If we push this picture, we  must note that since the Schwarzschild black hole is a non-extremal object, we are really dealing with a bound state, created by the  attractive gravitational force among the $N$ graviton
    particles.
   This is in contrast to the BPS branes, which are just coincident, since the gravitational and the $p$-form forces among the branes are opposite and cancel each other. 
   
   This picture finally raises the question, if there exist a one-dimensional
  dual conformal theory  living on the holographic screen of Rindler space, namely the one-dimensional Rindler cone of $M^{1,1}$, in analogy to the $SU(N)$ gauge theory on the boundary of $AdS_5$. This conformal theory should be closely related to the bound states of the $N$-gravitons at the quantum critical point, which is also expected to be described by sort of a  conformal theory of gapless modes.

\end{document}